\documentclass[useAMS,usenatbib]{mn2e}
\usepackage{graphics}
\usepackage{graphicx}
\usepackage{epsfig}
\usepackage{amssymb,amsmath}
\usepackage{times}
\usepackage{natbib}
\usepackage{subfigure}

\title[Ram pressure profiles in groups and clusters]{Ram pressure
  profiles in galaxy groups and clusters}

\author[T. E. Tecce et al.]{Tom\'as
  E. Tecce$^{1,2}$\thanks{E-mail: tomas@iafe.uba.ar}, Sof\'ia
  A. Cora$^{2,3}$ and Patricia B. Tissera$^{1,2}$\\
$^{1}$Instituto de Astronom\'ia y F\'isica del Espacio, C.C. 67 Suc. 28, 
  C1428ZAA Ciudad de Buenos Aires, Argentina\\
$^{2}$Consejo Nacional de Investigaciones Cient\'ificas y T\'ecnicas, 
  Rivadavia 1917, C1033AAJ Ciudad de Buenos Aires, Argentina\\
$^{3}$Facultad de Ciencias Astron\'omicas y Geof\'isicas, Universidad 
  Nacional de La Plata, and Instituto de Astrof\'isica de La Plata,
  Observatorio Astron\'omico,\\~Paseo del Bosque S/N, B1900FWA La Plata,
  Argentina}

\def \rvir{$r_\text{200}$}
\def \mvir{$M_\text{200}$}

\def \sagrp{{\scriptsize SAGRP}}
\def \sagrpa{{\scriptsize SAGRP-A}}
\def \sagrpf{{\scriptsize SAGRP-F}}

\begin{document}

\date{Accepted. Received; in original form}

\pagerange{\pageref{firstpage}--\pageref{lastpage}} \pubyear{2011}

\maketitle

\label{firstpage}

\begin{abstract}
Using a hybrid method which combines non-radiative hydrodynamical
simulations with a semi-analytic model of galaxy formation, we
determine the ram pressure as a function of halocentric distance
experienced by galaxies in haloes with virial masses 12.5~$\leq \log
(M_\text{200}h/M_\odot) <$~15.35, for redshifts 0~$\leq z \leq$~3. The
ram pressure is calculated with a self-consistent method which uses
the simulation gas particles to obtain the properties of the
intergalactic medium. The ram pressure profiles obtained can be well
described by beta profile models, with parameters that depend on
redshift and halo virial mass in a simple fashion. The fitting
formulae provided here will prove useful to include ram pressure
effects into semi-analytic models based on methods which lack gas
physics, such as dark matter-only simulations or the Press-Schechter
formalism.
\end{abstract}


\section[]{Introduction}
In the local Universe, the properties of galaxies depend on the
environment in which they reside. The star formation rates of
galaxies in denser environments such as galaxy clusters are lower than
for similar field galaxies (\citealt{kauff2004};
\citealt*{verdugo2008}) and the fraction of red galaxies at a given
stellar mass increases with environmental density
(\citealt{baldry2006}; \citealt*{martinez2008}). Disc galaxies in
clusters are deficient in H{\small I} when compared to similar
galaxies in the field. The deficiency increases towards the cluster
centre, and H{\small I}-deficient spirals are observed to have
truncated gas discs \citep[see][and references
  therein]{boselli2006}. These observations suggest that galaxies are
transformed from blue star-forming systems into red, passive ones via
physical processes that remove gas, thus suppressing the star
formation in the affected galaxies. 

One possible mechanism is ram pressure stripping (RPS) of galactic gas
caused by the interaction with the hot, high-temperature gas of
the intragroup or intracluster medium (ICM). The galaxy loses gas if
the ram pressure (RP) exerted by the ICM, $P_\text{ram} \equiv
\rho_\text{ICM} v^2$, where $v$ is the velocity of the galaxy relative
to the ICM, exceeds the gravitational restoring force of the galaxy
(\citealt{gg72}; for a recent review of observations and simulations
of RPS in individual galaxies see \citealt{roediger2009}). Prior to
affecting the cold gas discs, RPS could also be responsible for the
removal of the hot gas haloes of galaxies after they become satellites
of a cluster \citep{mccarthy2008,bekki2009}.

Recent results show that RPS is not a process exclusive to cluster
environments, and it can play a role in smaller systems such as galaxy
groups (e.g. \citealt*{rasmussen2006}; \citealt{rasmussen2008}) or for
dwarf satellites of giant galaxies (e.g. \citealt{mc2007};
\citealt*{mastropietro2009}). Thus, a significant number of cluster
galaxies could have been `pre-processed' in the smaller systems in
which they resided prior to infall into the cluster
\citep[e.g.][]{fujita2004,cortese2006}.

Semi-analytic modelling of galaxy formation is a powerful approach to
explore the effects of environment on galaxy properties. Such models
do not require heavy computing power, and allow the exploration of a
large parameter space at a fraction of the computational cost of fully
self-consistent simulations. Semi-analytic models have proven very
successful at reproducing several observed properties of galaxies such
as the local luminosity function \citep[see the review
  by][]{baugh2006}. RPS, however, was considered in such models only
in a few cases (\citealt{on2003}; \citealt{lanzoni2005};
\citealt{bdl08}; Font et al. \citeyear{font2008}). These models are
based on dark matter (DM)-only simulations, and thus have to resort to
analytical approximations to describe the ICM.

An alternative approach is presented by \citet[][hereafter T10]{rp1},
who combine a semi-analytic model with N-body/hydrodynamical
simulations. In this hybrid approach, the kinematical and
thermodynamical properties of the ICM are provided by the gas
particles of the hydrodynamical simulations. This results in a
self-consistent method which does not introduce additional free
parameters into the model, automatically taking into account local
variations of the density or the velocity field. T10 find that,
compared to their method, the use of analytic approximations results
in overestimations of the RP larger than 50 per cent for $z
>$~0.5. The disadvantage of the T10 approach is that it cannot be
combined with DM-only simulations, such as the widely used Millennium
simulation (Springel et al. \citeyear{millennium}), or with models
which use the Extended Press-Schechter (EPS) formalism
\citep{bond1991,bower1991,lc1993} to determine the growth of DM haloes.

In this paper we determine fitting formulae for the RP exerted on galaxies
as a function of halocentric distance obtained by using the T10 method, in
simulated galaxy groups and clusters of different masses. We show that
these RP profiles can be well described using beta profile models
\citep{beta}, with parameters depending on redshift and halo virial
mass. The formulae provided here can then be used in semi-analytic models
which lack the gas physics required by the T10 approach.

\section[]{Construction of the ram pressure profiles}
To calculate the RP exerted by the intergalactic medium on satellite
galaxies in groups and clusters at different redshifts, we use the
hybrid model {\scriptsize SAGRP} (see T10) which combines non-radiative
N-body/smoothed particle hydrodynamics (SPH) simulations of galaxy
clusters \citep{dolag05,dolag2009} with a semi-analytic model which handles
other baryonic physical processes such as gas cooling, star formation, and
feedback by supernovae and active galactic nuclei. The SPH simulations
are resimulations of the regions around five galaxy clusters with mass
$\sim$~10$^{14}\, h^{-1} M_\odot$ (labeled g1542, g3344,
  g6212, g676 and g914 in \citealt{dolag05,dolag2009}) and three
clusters with mass $\sim$~10$^{15}\, h^{-1} M_\odot$ (g51, g1
  and g8 in \citeauthor{dolag05}). These correspond to a $\Lambda$CDM
cosmology with $\Omega_m$~=~0.3, $\Omega_\Lambda$~=~0.7, Hubble
constant $h$~=~0.7 (in units of 100~km~s$^{-1}$~Mpc$^{-1}$), a baryon
density $\Omega_b$~=~0.039 and a power spectrum normalisation
$\sigma_8$~=~0.9. The DM particle mass is 1.13~$\times$~10$^9 \,h^{-1}
M_\odot$ and the gas particle mass 1.69~$\times$~10$^8 \,h^{-1} M_\odot$
\citep[for more details see][]{dolag2009}.

For each DM halo in the simulations, identified by means of a
friends-of-friends algorithm \citep[{\scriptsize FOF},][]{davis85}, we
search for all galaxies within the halo virial radius \rvir, defined as the
radius within which the mean mass density is 200$\rho_\text{crit}$. We use
these galaxies as tracers of the RP at their current position, calculating
for each one the RP they experience by using the properties of the
surrounding gas particles (see T10 for details). We select all haloes
with no contamination by boundary particles and with $\log
M_\text{200} \geq$~12.5, where $M_\text{200}$ is the total
mass within \rvir~(in units of $h^{-1} M_\odot$). In smaller haloes
the RP is very low, affecting only the smallest dwarf galaxies, and so
these haloes will not be considered in this analysis. 

We identify haloes in the 67 simulation snapshots available in the
redshift range 0~$\leq z \leq$~3. The total number of selected haloes
grows steadily from $N_\text{halo}$~=~29 at $z$~=~3 until it reaches a
peak of 155 haloes at $z \simeq$~0.5, decreasing afterwards to a final
value of $N_\text{halo}$~=~114 at $z$~=~0. The smallest {\scriptsize
  FOF} haloes selected are resolved with at least~2800 DM particles,
and a similar number of gas particles. Haloes with $\log M_\text{200}
\simeq$~14 are resolved with about 9~$\times$~10$^4$~DM particles, and
the most massive haloes are resolved with 10$^6$ DM particles or more.

\begin{figure}
  \centering
  \epsfig{file=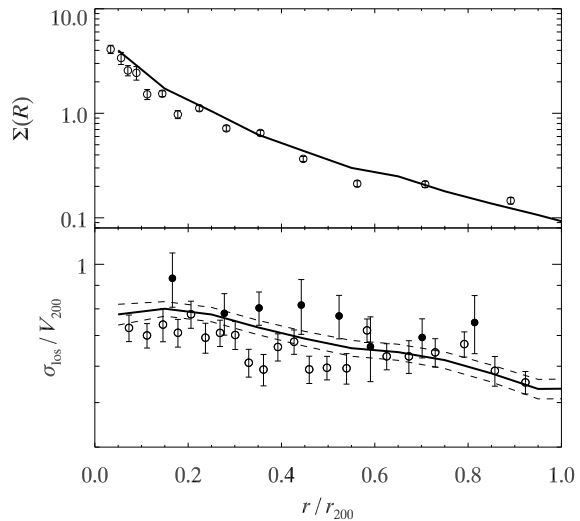, width=0.45\textwidth}
  \caption{Top: mean projected galaxy number surface density
      profile for model galaxies with $M_V < -$17 in all haloes with
      \mvir~$\geq$~10$^{14}\, h^{-1} M_\odot$ at $z$~=~0.3 (solid
      line). Open circles represent the mean observed surface density
      profile for CNOC survey clusters \citep{carlberg97a}. Bottom:
      mean line-of-sight velocity dispersion profiles for the same
      haloes, at $z$~=~0. Filled and open circles represent,
      respectively, data for galaxies with and without emission lines
      in the ENACS clusters, from \citet{bp2009}.}
  \label{fig:posvel}
\end{figure}

Positions and velocities of satellite galaxies within a
{\scriptsize FOF} halo are determined by the position and velocity
of the most bound DM particle of the corresponding subhalo (as done
also by \citealt{bdl08}). For those satellites whose subhalo has
been completely disrupted by the tidal forces exerted by the
gravitational potential of the main halo, we use the most bound
particle identified at the last time there was a subhalo. This is
essentially equivalent to assuming that galaxies trace the DM. In the
case of galaxy clusters, this assumption has been shown to provide a
good fit to observed galaxy number profiles out to large
clustercentric radii \citep[see e.g.][]{bg2003,gao2004}.

Using DM particles as tracers for galaxies allows us to
generate a population of cluster galaxies whose spatial and velocity
distributions match the observed ones. The top panel in
Fig.~\ref{fig:posvel} shows the galaxy number surface density
profile of simulated galaxies with absolute magnitude $M_V < -$17 in
all haloes with virial mass~$\geq$~10$^{14}\, h^{-1} M_\odot$ at
$z$~=~0.3 (solid line), compared with the observed profile for
cluster galaxies in the CNOC survey \citep*[][empty
  circles]{carlberg97a}. The density profiles for the simulations
are obtained by projecting along the $x$, $y$ and $z$ axes in turn,
considering only galaxies within 2\rvir~of the cluster centre,
binning the galaxies out to 2\rvir~and averaging over the three
projections \citep[as in][]{gao2004}. The mean galaxy surface density
profile thus obtained agrees very well with the observational
data.

The velocity dispersions that we obtain from galaxies
contained within massive haloes ($\sigma \sim$~400~km~s$^{-1}$ for
haloes with $\log M_\text{200} \simeq$~14 and $\sigma
\sim$~900~km~s$^{-1}$ for haloes with $\log M_\text{200} \simeq$~15)
are consistent with the mass-velocity dispersion scaling relations
determined from observations \citep[e.g.][]{hl2008,wl2010}. The
bottom panel of Fig.~\ref{fig:posvel} shows the mean line-of-sight
velocity dispersion profile of the simulated galaxies at $z$~=~0
(solid line), constructed in the same way as the density profiles;
dashed lines show the 1$\sigma$ errors. Circles show the data for
ENACS clusters from \citet{bp2009}, where filled and open symbols
represent data for galaxies with and without emission lines,
respectively. Although our model galaxies do not present the orbital
dichotomy between early and late type galaxies observed (see
e.g. \citealt{sodre89}; \citealt*{adami98}; \citealt{bk2004}), the
mean velocity dispersion profile is a very good match to the average
observational trend. Therefore, we find that our choice of galaxy
orbits is consistent with cluster observations, and we make the
assumption that this method will provide a good approximation to the
orbits of galaxies in less massive haloes as well.

In T10 it is shown that the mean RP inside a halo increases with redshift
and cluster mass. Therefore, we split all the selected haloes in 10
logarithmic mass intervals in the range 12.5~$\leq \log M_\text{200}
<$~15.35. The adopted bin width is $\Delta \log
M_\text{200}$~=~0.5. The bins overlap by half an interval,
i.e. 12.5~$\leq \log M_\text{200} <$~13; 12.75~$\leq \log M_\text{200}
<$~13.25 and so on. By doing this we have additional points for
fitting the mass dependence, in a procedure which resembles a
(short-period) moving average. For each mass bin we construct a
combined profile of RP versus halocentric distance by gathering the
data of all the galaxies within \rvir~of the corresponding haloes. We
scale the halocentric galaxy distances by the \rvir~of their host
halo, and we determine the median RP in equally-spaced radial intervals.
 
\begin{figure}
  \centering
  \epsfig{file=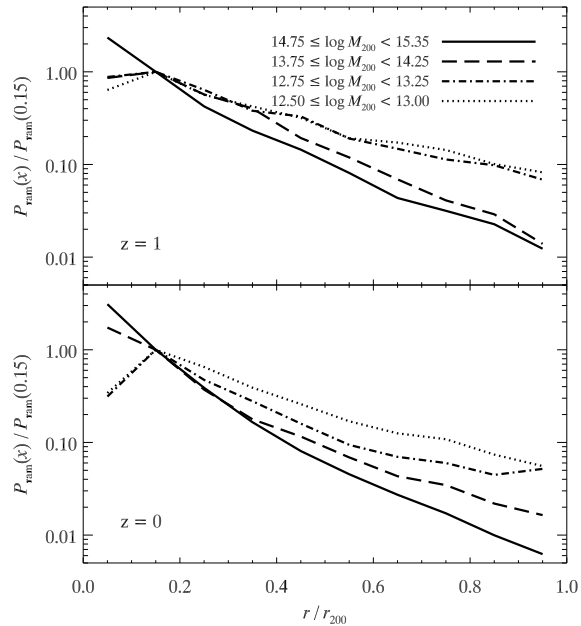, width=0.45\textwidth}
  \caption{Combined RP profiles for haloes in four selected virial
    mass ranges at $z$~=~1 (top) and $z$~=~0 (bottom). Each profile is
    normalised to its value at $r_N / r_\text{200}$~= 0.15, which for the
    less massive haloes is equivalent to 6 times the gravitational
    softening used in the N-body/hydrodynamical simulations.}
  \label{fig:rpprof}
\end{figure}

To determine whether the shapes of the RP profiles obtained depend on
virial mass and redshift, we renormalise them to their value at $x_N
\equiv r_N / r_\text{200}$~=~0.15. For the less massive haloes this is
equivalent to 6 times the gravitational softening used in the SPH
simulations. As an example, Fig.~\ref{fig:rpprof} shows the combined
RP profiles determined for four different mass bins at $z$~=~1 (top
panel) and $z$~=~0 (bottom panel). As can be seen, the combined RP
profiles show different slopes for different mass intervals and
times. The RP gradient is stronger in the more massive clusters, where
the RP at $r$~=~\rvir~ is $\sim$1 per cent of the value at $x_N$; in
the less massive haloes, the RP at \rvir~is~$\sim$10 per cent of the
central value. We also find, at a given halo mass, a dependence on
redshift of the normalisation of the RP profiles. For example, the
average RP value at $x_N$ varies from $\sim$1.8~$\times$~10$^{-10}\,
h^2$~dyn~cm$^{-2}$ at $z$~=~1 to $\sim$4.7~$\times$~10$^{-10}\,
h^2$~dyn~cm$^{-2}$ at $z$~=~0 for haloes with 14.75~$\leq \log
M_\text{200} <$~15.35, and from $\sim$2.1~$\times$~10$^{-12}\,
h^2$~dyn~cm$^{-2}$ at $z$~=~1 to $\sim$4.6~$\times$~10$^{-12}\,
h^2$~dyn~cm$^{-2}$ at $z$~=~0 for haloes with 12.5~$\leq \log
M_\text{200} <$~13. These trends suggest the need to adopt a fitting
function with at least two parameters, normalisation and slope, both
depending on halo virial mass and redshift.

\section[]{Fits to the profiles}
To fit the combined RP profiles obtained at each simulation snapshot for
the different halo mass ranges selected, three different analytical
functions are considered: a simple power law,
\begin{equation}\label{eq:powerlaw}
  P_\text{ram} \propto (r /r_s^P)^{-\alpha}, 
\end{equation}
a \citet*[][hereafter NFW]{nfw} profile, which
provides a good description of the density profiles of DM haloes,
\begin{equation}\label{eq:nfw}
  P_\text{ram} \propto (r / r_s^N)^{-1} \left[1+ (r / r_s^N) \right]^{-2},
\end{equation}
and a beta profile, which is commonly used to fit the density profiles
of the ICM in galaxy clusters, 
\begin{equation}\label{eq:beta}
  P_\text{ram} \propto \left[1 + (r / r_s^B)^2 \right]^{-3\beta/2}.
\end{equation}
In the above relations $r_s^P$, $r_s^N$ and $r_s^B$ are the corresponding
characteristic radii. When used to describe ICM profiles, the beta model is
usually chosen to have a fixed exponent $\beta$~=~2$/$3. In our case, we
fit both a model with this fixed value,  and another one where $\beta$ is
allowed to vary; these models will be hereafter called fixed beta and
full beta model, respectively.

\begin{figure}
  \centering
  \epsfig{file=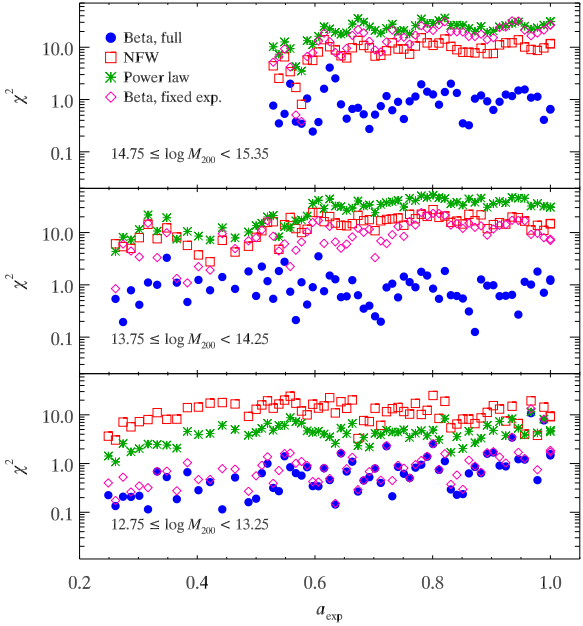, width=0.45\textwidth}
  \caption{Chi-square goodness-of-fit estimator as a function of
    cosmological expansion factor $a_\text{exp}$ for the four analytic
    profile models considered, and for three selected mass ranges:
    14.75~$\leq \log M_\text{200} <$~15.35 (top), 13.75~$\leq \log
    M_\text{200} <$~14.25 (centre) and 12.75~$\leq \log M_\text{200}
    <$~13.25 (bottom). In each mass range, $\chi^2$ values for all
    models are scaled to the average value obtained for the full beta
    model.}
  \label{fig:chi}
\end{figure}

We determine the best-fitting profile of the
form~\eqref{eq:powerlaw},~\eqref{eq:nfw} and~\eqref{eq:beta} (in this
latter case, for both the fixed and the full beta models) for the combined
RP profiles in the ten mass intervals considered and as a function of
redshift. All fittings are carried out using the Levenberg-Marquardt
technique to solve the least-squares problem \citep{mpfit}. For each fit,
we compute the chi-square goodness-of-fit estimator. The chi-square values
are plotted in Fig.~\ref{fig:chi} as a function of the cosmological
expansion factor $a_\text{exp}$ of the corresponding simulation snapshots,
for three mass ranges selected for illustration purposes only. To make the
comparison easier, the chi-square values in Fig.~\ref{fig:chi} have all
been normalised to the average chi-square value obtained for the full beta
model, which as can be clearly seen from the figure, provides the
best-fitting profile for all times and mass ranges. Therefore, we propose
as a model for the RP profiles determined with the T10 method a full beta
model~\eqref{eq:beta} with a core value $P_0$, scale radius $r_s$ and
exponent $\beta$ all depending on both virial mass and redshift,
\begin{equation}\label{eq:rpbeta}
  P_\text{ram}(M,z) = P_0(M,z) \left[1 +
    \left(\frac{r}{r_s(M,z)}\right)^2
    \right]^{-\frac{3}{2}\beta(M,z)}.
\end{equation}

\begin{figure}
  \centering
  \epsfig{file=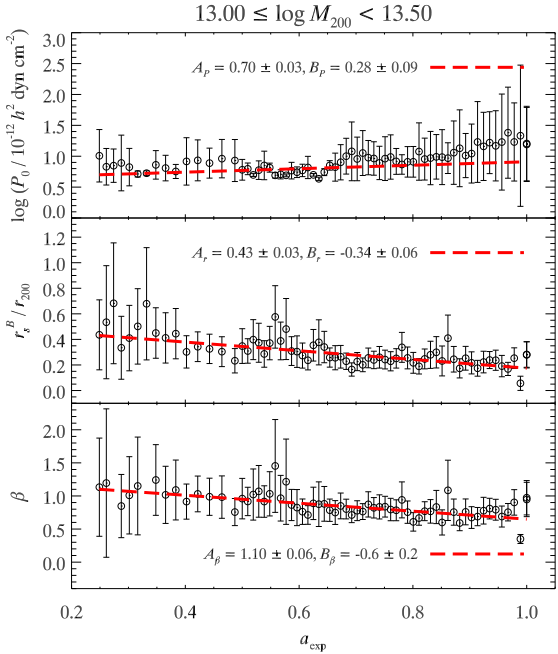, width=0.45\textwidth}
  \caption{Evolution with time of the best-fitting parameters for the
    full beta model, for the mass range 13~$\leq \log M_\text{200} <$~13.5
    selected as an example: logarithm of the core value $P_0$ (top),
    scale radius $r_s / r_\text{200}$ (centre) and exponent~$\beta$
    (bottom). The dashed lines indicate linear fits to the corresponding
    parameters, of the form $A + B (a_\text{exp}-0.25)$.}
  \label{fig:fit1}
\end{figure}

The next step is to determine how the model parameters in
\eqref{eq:rpbeta} depend on DM halo mass and redshift. To do this we plot
for each mass interval the best-fitting parameters as a function
of $a_\text{exp}$. We find that for all the mass intervals considered, the
evolution of the best-fitting parameters can be well fitted with a linear
regression. As an example, Fig.~\ref{fig:fit1} shows the evolution with
time of the best-fitting values of $\log P_0$, $r_s$ and $\beta$ obtained
for haloes in the range 13~$\leq \log M_\text{200} <$~13.5.

The zero point and slope of the linear fits change for the different
mass ranges. Hence, we propose a dependence of the model parameters of
the following form: 
\begin{subequations}\label{eq:params}
  \begin{align}
    \log \left( \frac{P_0}{10^{-12} h^2 \text{dyn cm}^{-2}} \right) & =
    A_P + B_P (a_\text{exp} - 0.25),\\
    \frac{r_s}{r_\text{200}} & = A_r + B_r (a_\text{exp}-0.25),\\
    \beta & = A_\beta + B_\beta (a_\text{exp}-0.25),
  \end{align}
\end{subequations}
where in the above relations, the coefficients $A$ and $B$ all depend, in
principle, on halo virial mass. We have chosen to express the RP values in
units of 10$^{-12}\, h^2$~dyn~cm$^{-2}$, which is of the order of the core
value of RP in the smallest haloes considered. 

\begin{figure*}
  \centering
  \subfigure{
    \epsfig{file=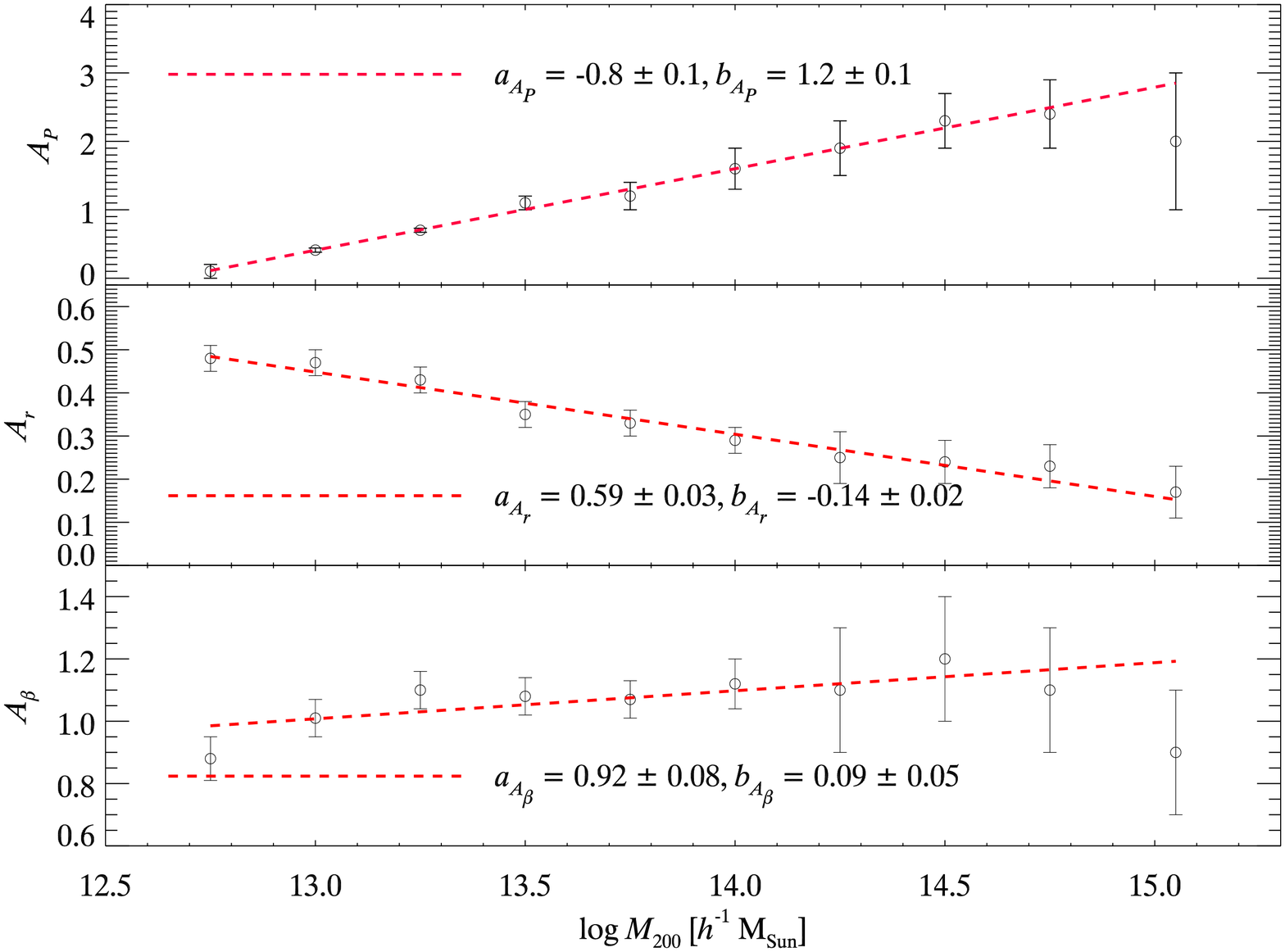, width=0.45\textwidth}
  }
  \subfigure{
    \epsfig{file=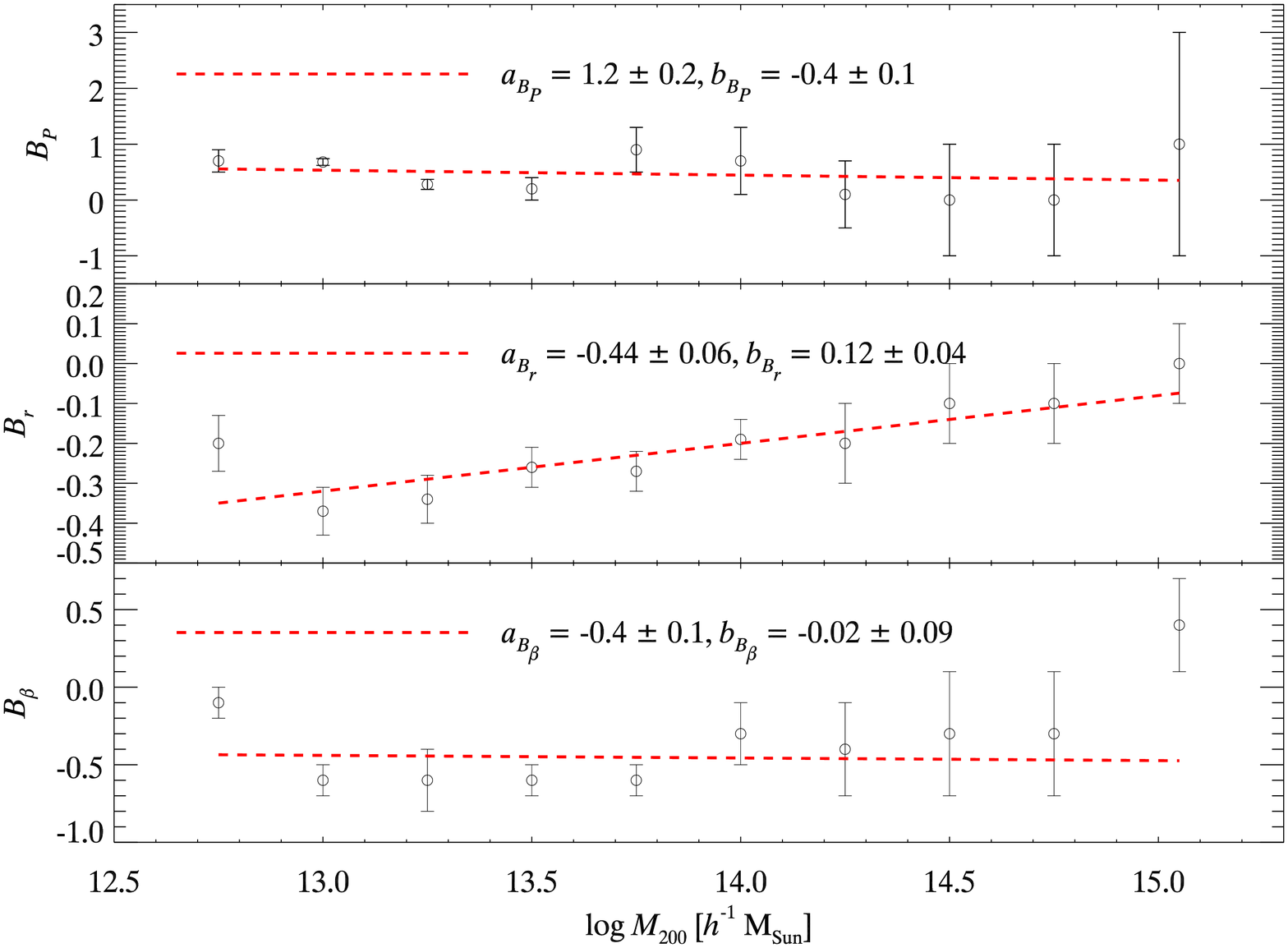, width=0.45\textwidth}
  }
  \caption{Values of the parameters obtained when fitting the time
    evolution of the best-fitting parameters for the full beta model
    \eqref{eq:rpbeta} with a straight line $A + B (a_\text{exp}-0.25)$,
    plotted as a function of halo virial mass: $\log P_0$ (top panels),
    $r_s/r_\text{200}$ (centre) and $\beta$ (bottom). Left and right
    panels show the corresponding values for the coefficients $A$ and
    $B$, respectively. In all cases, the mass dependence of the
    coefficients can be fitted again with a linear regression of the
    form $a + b(\log M_\text{200} - 12.0)$, shown with a dashed line.}
  \label{fig:fit2}
\end{figure*}

The fitting coefficients for the linear regressions \eqref{eq:params}
corresponding to each virial mass range are shown in
Fig.~\ref{fig:fit2}. In all cases, the dependence of the linear fit
coefficients $A$ and $B$ on virial mass can again be very well fitted by a
linear regression, chosen to be of the form $a + b (\log M_\text{200}
- $12$)$. Combining the mass and redshift dependences, we finally
obtain the following expressions for the coefficients in
\eqref{eq:params}: 
\begin{subequations}\label{eq:paramfits}
\begin{align}
  A_P & = (-0.8 \pm 0.1) + (1.2 \pm 0.1)(\log M_\text{200} - 12), \\
  B_P & = (1.2 \pm 0.2) + (-0.4 \pm 0.1)(\log M_\text{200} - 12), \\
  A_r & = (0.59 \pm 0.03) + (-0.14 \pm 0.02)(\log M_\text{200} - 12), \\
  B_r & = (-0.44 \pm 0.06) + (0.12 \pm 0.04)(\log M_\text{200} - 12), \\
  A_\beta & = 0.92 \pm 0.08,\label{eq:abeta}\\
  B_\beta & = -0.4 \pm 0.1.\label{eq:bbeta}
\end{align}
\end{subequations}

To illustrate the potential of our model, we have run a modified
version of \sagrp~in which we replaced the RP calculation using the
gas particles from the underlying simulations by the fitting formulae
determined in this work. We will hereafter refer to this modified model as
\sagrpf. In \sagrpf, as an approximation we set the RP equal to zero for
all galaxies if $z >$~3, and for $z$~=~3 onwards we determine, for all
satellite galaxies in each halo, the distance to the halo central galaxy
and then the corresponding RP using~\eqref{eq:rpbeta}, \eqref{eq:params}
and~\eqref{eq:paramfits}. We compare the results of \sagrpf~with those from
\sagrp~and also with \sagrpa, a model where the RP is calculated
analytically by using a NFW profile for the ICM density and assuming the
ICM to be in hydrostatical equilibrium (see T10 for details).

\begin{figure*}
    \epsfig{file=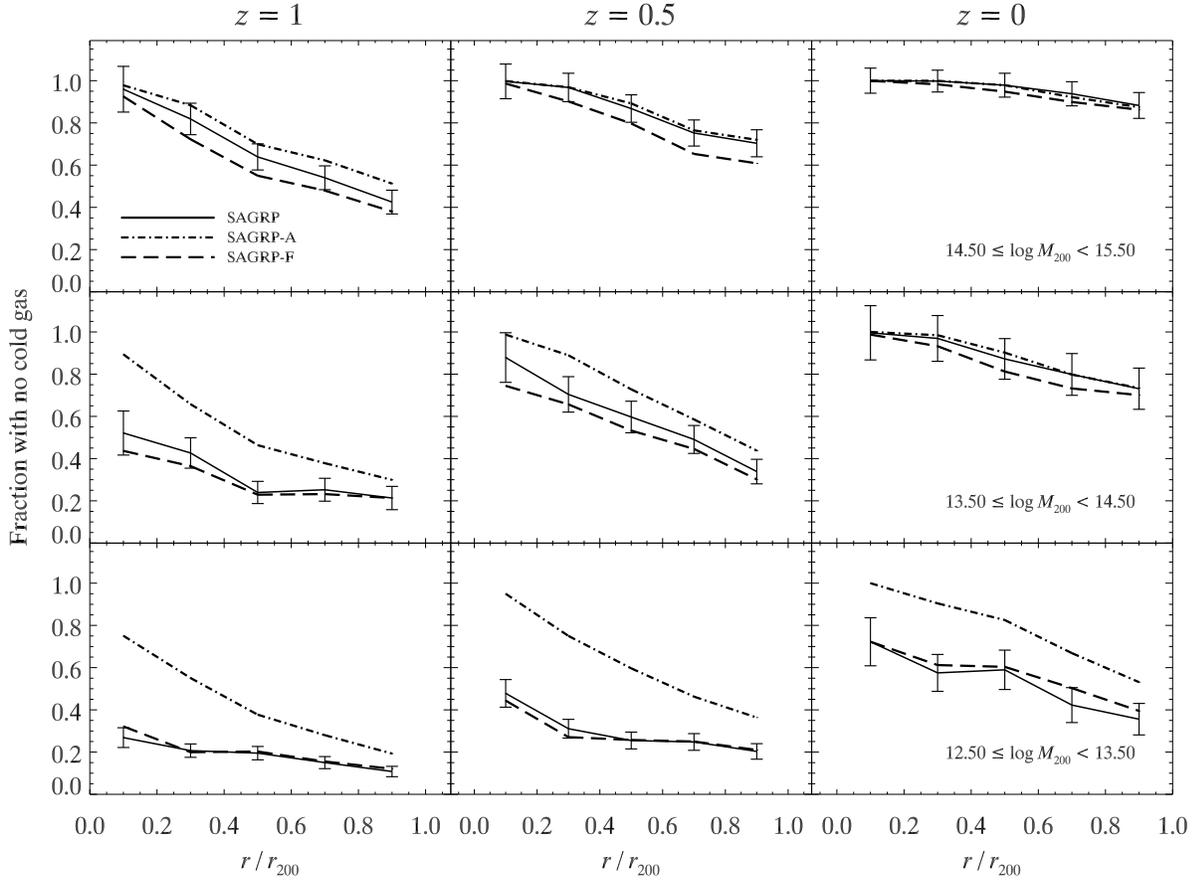, width=0.9\textwidth}
  \caption{Fraction of galaxies with stellar mass $\geq$~10$^9\,
    h^{-1} M_\odot$ that have completely lost their cold gas as a
    function of halocentric distance, for three different models:
    \sagrp, which determines the RP from the information provided by
    the gas particles in the underlying simulations (solid lines),
    \sagrpa, which calculates the RP by using an analytical
    approximation (dot-dashed lines), and \sagrpf, which uses the
    best-fitting formulae determined in this work (dashed
    lines). Results are presented for $z$~=~1, 0.5 and 0 (left, centre
    and right columns, respectively) and for haloes in three different
    mass ranges: 14.5~$\leq \log M_\text{200} \leq$~15.5 (top row),
    13.5~$\leq \log M_\text{200} <$~14.5 (centre row) and 12.5~$\leq
    \log M_\text{200} <$~13.5 (bottom row). Error bars show the
    1$\sigma$ errors for the \sagrp~model.}
  \label{fig:fgas}
\end{figure*}

Fig.~\ref{fig:fgas} shows the fraction of satellite galaxies with
stellar mass~$\geq$~10$^{9}\, h^{-1} M_\odot$ that have completely
lost their cold gas mass as a function of halocentric distance, for the
three different models runs: \sagrp~(solid lines), \sagrpa~(dot-dashed
lines), and \sagrpf~(dashed lines). Results are shown grouping the halo
masses in three separate bins: 14.5~$\leq \log M_\text{200}
\leq$~15.5, 13.5~$\leq \log M_\text{200} <$~14.5 and 12.5~$\leq \log
M_\text{200} <$~13.5 (top, centre and bottom rows, respectively), and
for three selected redshifts: $z$~=~1 (left column), $z$~=~0.5 (centre
column) and $z$~=~0 (right column).

As already shown in T10, calculating the RP by using an analytical
approximation is a good match for the most massive clusters at $z \sim$~0,
but grossly overestimates the RP for $\log M_\text{200} \lesssim$~14
at $z >$~0. On the other hand, Fig.~\ref{fig:fgas} clearly shows that
our best-fitting formulae provide an excellent match to the results of
the self-consistent RP estimation of \sagrp, matching the slope and
evolution of the fractions in all mass ranges and at all the redshifts
considered. For the two most massive ranges considered (top and centre
rows in Fig.~\ref{fig:fgas}), the \sagrpf~model appears to
systematically underestimate the RP; however, the diference is within
the 1$\sigma$ errors for the models.

\section[]{Conclusions}
Using a hybrid method that combines a semi-analytic model of galaxy
formation with cosmological hydrodynamical simulations, which takes
advantage of the extra kinematical and thermodynamical information
provided by the simulation gas particles, we have determined the RP as
a function of halocentric distance for DM haloes with virial
masses 12.5~$\leq \log M_\text{200} <$~15.35, for redshifts
in the range 0~$\leq z \leq$~3. The RP profiles found can be well
described by beta models, with parameters that depend on virial mass
and redshift in a relatively simple fashion. The resulting
prescriptions can be used to include the effect of RPS in
semi-analytic models in which the growth of DM haloes is determined
either from cosmological N-body simulations which do not include gas
particles, or with an approach based in the EPS formalism
(e.g. \citealt{st2002}; \citealt{cole2008}; \citealt*{neistein2010}).

In our best-fitting formulae for the RP profiles given
by~\eqref{eq:rpbeta}, \eqref{eq:params} and~\eqref{eq:paramfits}, we
note that whereas the core RP value $P_0$ and scale radius $r_s$
depend on both halo virial mass and redshift, the exponent $\beta$
depends only on time. We have checked this by running several
different variants of the \sagrpf~model, including in equations
\eqref{eq:abeta} and \eqref{eq:bbeta} a mass-dependent term similar to
those in the relations for the other parameters, varying the
coefficients $b_{A_\beta}$ and $b_{B_\beta}$ (see Fig.~\ref{fig:fit2})
within the range of errors obtained for them. In all cases, we find
that the smallest difference between \sagrp~and \sagrpf~is obtained
when~$\beta$ is assumed to be independent of virial mass.

The results presented in this work extend those of T10 and
\citet{bdl08}, who focused on the most massive clusters with $\log
M_\text{200} \gtrsim$~14, by also considering the RP profiles in less
massive galaxy groups. The gradient of RP becomes steeper as the
virial mass increases; in the massive clusters, the RP at the core is
$\sim$100 times higher than at $r =$~\rvir, whereas in galaxy
group-sized haloes the RP at \rvir~is about 10 per cent of the central
value. On the other hand, the RP in the outskirts of $\log
M_\text{200}$~=~15 clusters is $\sim$5~$\times$~10$^{-12}\,
h^{2}$~dyn cm$^{-2}$ at $z$~=~0, of the same order of magnitude as the
RP experienced by galaxies in the cores of haloes with $\log
M_\text{200}$~=~13.5 at the same epoch. Using SPH simulations,
\citet{roediger2006} find that such levels of RP can remove about a
quarter of the total gas of a spiral galaxy with mass
$\sim$2~$\times$~10$^{11}\, M_\odot$. One concludes that dwarf
galaxies with stellar mass~$\lesssim$~10$^{9}\, M_\odot$ in groups of
all masses are very likely to experience gas loss by RPS, at least in
the local Universe when the RP levels are higher in all haloes.

The fits obtained in this work were determined by using a set of
resimulations including gas physics of the regions surrounding eight
massive galaxy clusters, extracted originally from a DM-only
cosmological simulation. These regions are large enough to contain a
fair sample of haloes in the mass range considered, free of
contamination from boundary particles. Haloes in the lowest
mass bin are resolved with at least 2800 DM particles, and the most
massive haloes are resolved with 10$^6$ DM particles or more.
However, we are restricted to only a few massive clusters (there are
only three haloes in the most massive bin), and this increases the
relative error in the fitting parameters. 

The RPS depends on the orbits of the satellite galaxies. There
are several different approaches to determining the orbits of
satellites in a semi-analytic model (see
e.g. \citealt{lanzoni2005,cora2006}; Font et
al.~\citeyear{font2008}). As in \citet{bdl08}, we track the orbital
evolution of satellite galaxies within haloes by following the
most-bound DM particle of their subhalo. We have shown that the
method chosen results in positions and velocities of cluster
galaxies which are in good agreement with the mean observed trends,
although in the case of the velocity dispersion profiles the
dichotomy between early and late type galaxies is not seen in the
model galaxies. This may point to missing physics in the models,
since it is still unclear which process is responsible for the
observed differences \citep[see][]{bp2009}.

The particular profile fits obtained could also depend on the
method chosen for the hydrodynamical calculation in the
simulations. For example, the non-radiative SPH simulations used in
this work may suffer from artificially suppressed turbulence (see
\citealt{dolag05} and Agertz et al. \citeyear{agertz2007}). Using
simulations carried out using a grid-based method, for example,
could result in increased turbulence in the ICM and lead to larger
scatter in the profiles, as already noted in T10. We intend to
explore this issue in the future, as larger simulations including
different implementations for the gas physics become available to
us.

Notwithstanding the above considerations, within the current knowledge
and taking into account the commonly used hypotheses, our best-fitting
formulae capture remarkably well the effects of RP as given by the
hybrid approach of T10 on galaxies in different DM halo masses at
different redshifts, and provide a significant improvement over the
analytical approximations used so far to estimate RPS effects.

\section*{Acknowledgments}
The authors thank the referee for useful comments. They would also
like to thank Klaus Dolag for making the hydrodynamical simulations
available to them, and Ho Seong Hwang and Andrea Biviano for providing
observational data. This work was partially supported by PICT 245 Max
Planck (2006) and PIP 0305/09 (CONICET).

\bibliography{mn-jour,ttecce_rpprofiles_rev2}

\label{lastpage}

\end{document}